\documentclass[pra,twocolumn,showpacs,longbibliography]{revtex4-1}

\usepackage[latin1]{inputenc}
\usepackage{exscale}
\usepackage{graphicx}
\usepackage{amsmath}
\usepackage{amssymb}
\usepackage{color}

\newcommand{\ee}{\mathrm{e}}
\newcommand{\ii}{\mathrm{i}}
\newcommand{\dd}{\mathrm{d}}

\newcommand{\tr}{\mathcal{T}}
\newcommand{\va}{\varepsilon}

\newcommand{\an}[1]{\hat{#1}}
\newcommand{\cre}[1]{\hat{#1}^\dag}
\newcommand{\av}[1]{\langle #1\rangle}

\begin{document}

\title{Mesoscopic transport of fermions through an engineered optical lattice\\connecting two reservoirs}

\author{M. Bruderer}
\author{W. Belzig}

\affiliation{Fachbereich Physik, Universität Konstanz, D-78457 Konstanz, Germany}

\date{\today}

\begin{abstract}
We study transport of fermions in a system composed of a short optical
lattice connecting two finite atomic reservoirs at different filling levels.
The average equilibration current through the optical lattice, for strong
lattice-reservoir coupling and finite temperatures, is calculated within the
Landauer formalism using a nonequilibrium Green's functions approach. We
moreover determine quantum and thermal fluctuations in the transport and
find significant shot-to-shot deviations from the average equilibration
current. We show how to control the atomic current by engineering specific
optical lattice potentials without requiring site-by-site manipulations and
suggest the realization of a single level model. Based on this model we
discuss the blocking effect on the atomic current resulting from weak
interactions between the fermions.
\end{abstract}


% 05.60.Gg 	Quantum transport
% 67.85.-d 	Ultracold gases, trapped gases
% 67.85.Pq 	Mixtures of Bose and Fermi gases
% 72.10.-d 	Theory of electronic transport; scattering mechanisms

\maketitle

%%%%%%%%%%%%%%%%%%%%%%%%%%%%%%%%%%%%%%%%%%%%%%%%%%%%%%%%%%%%
%%%%%%%%%%%%%%%%%%%%%%%%%%%%%%%%%%%%%%%%%%%%%%%%%%%%%%%%%%%%
%%%%%%%%%%%%%%%%%%%%%%%%%%%%%%%%%%%%%%%%%%%%%%%%%%%%%%%%%%%%

\section{Introduction}

Ultracold atoms in optical lattices have been proven to be perfectly suited
for implementing physical models of interest to the field of atomic and
condensed matter physics~\cite{Bloch-RMP-2008,Esslinger-AR-2010}.
Specifically, an important part of the related experimental efforts have
improved and are still extending our understanding of nonequilibrium quantum
transport. These efforts have resulted in the observation of fundamentally
interesting quantum mechanical processes such as Bloch
oscillations~\cite{Roati-PRL-2004}, Landau-Zener
tunneling~\cite{Zenesini-AX-2007}, and interaction-controlled
transport~\cite{Strohmaier-PRL-2007}. Studying nonequilibrium transport in
optical lattices has several advantages over conventional condensed matter
systems: Ultracold atoms exhibit slow coherent quantum dynamics (with
kilohertz tunneling rates) and are detectable in small numbers on
microscopic scales. In addition, the ability to tune the interactions
between atoms via Feshbach resonances makes it possible to investigate
transport of interacting and noninteracting particles.

While many transport-related experiments employed tilted lattice potentials,
alternative setups for studying nonequilibrium phenomena have been
suggested, in which a current of particles flows between two atomic
reservoirs. Micheli~\emph{et
al.}~\cite{Micheli-PRL-2004,Micheli-PRA-2006,Daley-PRA-2005} pointed out
that, analogous to a transistor, an impurity atom may be utilized to control
the flow of a one-dimensional Bose or Fermi gas. They considered the full
time-dependent coherent dynamics of the ultracold gas in a closed system and
determined the particle current by using analytical
approximations~\cite{Micheli-PRA-2006} and time-dependent density matrix
renormalization group (DMRG) calculations~\cite{Daley-PRA-2005}.

Pepino~\emph{et~al.}~\cite{Pepino-PRL-2009,Pepino-PRA-2010} generalized this
idea by replacing the single impurity with an optical lattice coupled to
separate bosonic reservoirs in order to emulate the behavior of
semiconductor electronic circuits (see also \cite{Seaman-PRA-2007}). In
Refs.~\cite{Pepino-PRL-2009,Pepino-PRA-2010} reservoirs were introduced and
specified as large sources or sinks of particles at zero temperature with a
Fermi-sea-like energetic distribution, constant chemical potential, and
fast-decaying system-reservoir correlations. A quantum master equation,
relying on weak system-reservoir coupling, was used to describe time
evolution of the system, thereby in part eliminating the coherent evolution
of the reservoir.

\begin{figure}[h!]
\begin{center}
  \includegraphics[width = 240pt]{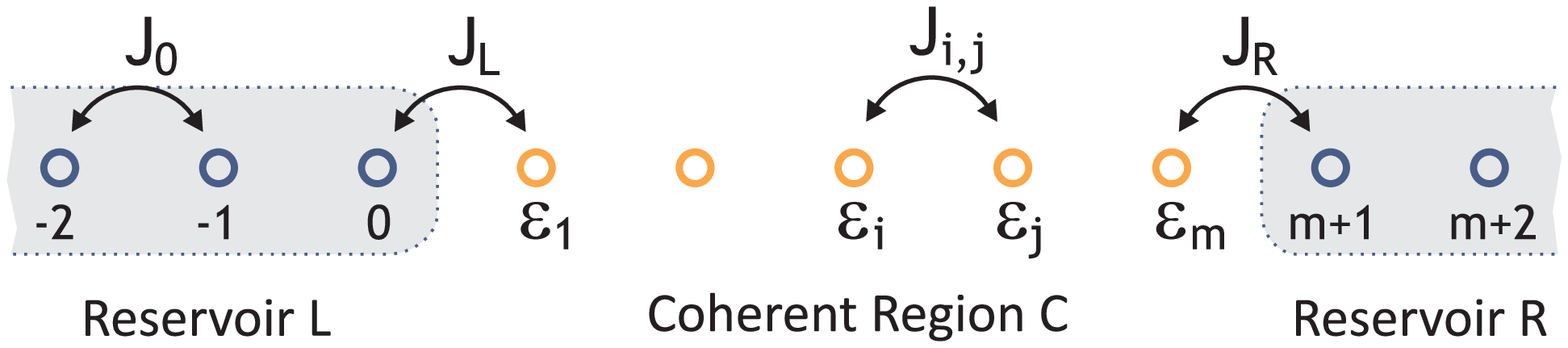}
  \caption{(Color online) Fermions confined to an optical lattice
  hop from the left reservoir $L$ through a short coherent part~$C$ (sites $1$ to $m$)
  into the right reservoir $R$. The hopping parameters $J_{i,j}$,
  the on-site energies $\va_j$, and the couplings $J_L$ and $J_R$ may
  take arbitrarily engineered values, whereas the hopping parameter in the
  reservoirs $J_0$ is held constant.}\label{scheme}
\end{center}
\end{figure}

Here we consider the evolution of a one-dimensional Fermi gas loaded into an
optical lattice which is partitioned into two large incoherent reservoirs
$L$ and $R$ connected by a short coherent region~$C$, as illustrated in
Fig.~\ref{scheme}. In this setup the difference in the chemical potentials
$\mu_L$ and $\mu_R$ of reservoirs $L$ and $R$ drives a current of fermionic
atoms through the coherent region $C$.

%%%%%%%%%%%%%%%%%%%%%%%%%%%%%%%%%%%%%%%%%%%%%%%%%%%%%%%%%%%%
%%%%%%%%%%%%%%%%%%%%%%%%%%%%%%%%%%%%%%%%%%%%%%%%%%%%%%%%%%%%

In contrast to previous theoretical works, we take a mesoscopic perspective
on nonequilibrium transport between the atomic reservoirs: We analyze the
evolution of the Fermi gas within the Landauer
formalism~\cite{Imry-RMP-1999,Nazarov-2009}, where transport is described as
a transmission process through the coherent region $C$ with fermions emitted
and absorbed by the reservoirs $L$ and $R$, respectively. In this vein, we
understand reservoirs to be finite containments filled with noninteracting
fermions in the ground state, characterized by a Fermi distribution with
finite temperature and well-defined (but not necessarily constant) chemical
potential. The transmission is determined by a nonequilibrium Green's
functions (NEGF) approach in the tight-binding
picture~\cite{Caroli-JPC-1971,Cuevas-2010}. This approach allows us to
consider a wide range of experimentally relevant parameters. Moreover, we
express our results in terms of the number of atoms accumulated in the
reservoirs, which (in particular for small currents) may be more accessible
in experiments than the current.

Accordingly, we complement and significantly extend the results in
Refs.~\cite{Pepino-PRL-2009,Pepino-PRA-2010} in several directions for the
case of fermionic atoms: First, we take into account that experimental
ultracold systems are finite and therefore we describe not only
instantaneous steady-state currents at constant chemical potentials but the
full equilibration process between the reservoirs. Second, our analysis is
valid for arbitrarily engineered configurations of the coherent region $C$
and for strong coupling to the reservoirs, thereby going beyond the
weak-coupling expansion in Refs.~\cite{Pepino-PRL-2009,Pepino-PRA-2010}.
Strong coupling indeed corresponds to the most elementary setup with the
parts $L$, $R$ and $C$ consisting of a single homogeneous optical lattice.
Third, by using the tools of full counting statistics~\cite{Nazarov-2009}
and considering intrinsic damping mechanisms we explicitly evaluate the
evolution of the fluctuations in the current. These fluctuations indicate
large shot-to-shot deviations from the average value in repeated
measurements of the number of particles in the reservoirs. Our approach
includes finite-temperature effects relevant to both the average current and
the fluctuations. For instance, thermal fluctuations are shown to build up
during the equilibration process until they reach a constant value
proportional to the temperature of the Fermi gas.

In the first part of this paper we analyze the evolution of the Fermi gas
within the NEGF-Landauer formalism. In this framework we discuss the
properties of the system in terms of the average current, the filling levels
of the reservoirs, and fluctuations of these quantities. In the second part
we first apply the formalism to the case of constant transmission between
the reservoirs and subsequently discuss more complex situations. In
particular, we show how to control the current by modulating the connecting
optical lattice with the help of additional optical potentials. Finally, we
suggest a way to realize a single-level model and demonstrate that in this
model even weak interactions between the fermions are sufficient to reduce
the atomic current.

%%%%%%%%%%%%%%%%%%%%%%%%%%%%%%%%%%%%%%%%%%%%%%%%%%%%%%%%%%%%
%%%%%%%%%%%%%%%%%%%%%%%%%%%%%%%%%%%%%%%%%%%%%%%%%%%%%%%%%%%%
%%%%%%%%%%%%%%%%%%%%%%%%%%%%%%%%%%%%%%%%%%%%%%%%%%%%%%%%%%%%

\section{NEGF-Landauer Model}

We start with the theoretical framework required to determine the average
equilibration current through the lattice and the filling levels of the
reservoirs. The Hamiltonian of the system within the lowest Bloch band and
in the tight-binding approximation is $\hat{H}_{C} + \hat{H}_{L} +
\hat{H}_{R} + \hat{H}_{I}$ with
\begin{equation}\label{ham}
\begin{split}
  &\hat{H}_{C} = -\sum_{\langle i,j\rangle}J_{i,j}\cre{c}_i\an{c}_j + \sum_{j} \va_j\cre{c}_j\an{c}_j\qquad i,j\in 1..m\\
  &\hat{H}_{L} = \hat{H}_{R} = -J_0\sum_{\langle i,j\rangle}\cre{c}_i\an{c}_j + \hat{H}_{S}\qquad i,j\notin 1..m\\
  &\hat{H}_{I} = -J_L(\cre{c}_1\an{c}_0 + \cre{c}_0\an{c}_1) - J_R(\cre{c}_m\an{c}_{m+1} + \cre{c}_{m+1}\an{c}_{m})\,,
\end{split}
\end{equation}
where the central part $C$ is formed by the sites $1$ to $m$ and $\langle
i,j\rangle$ denotes the sum over nearest neighbors. The operators
$\cre{c}_j$ ($\an{c}_j$) create (annihilate) a spin-polarized fermion in a
Wannier state localized at site $j$. The hopping parameters in the central
part and in the reservoirs are $J_{i,j}$ and $J_0$, respectively, and
$\va_j$ are on-site energies. The couplings $J_{L}$ and $J_{R}$ connect the
central part to the reservoirs which are each composed of $M$ sites.

The Hamiltonian $\hat{H}_S$, not specified explicitly, represents
interactions of the reservoirs with an engineered environment, e.g., with an
atomic gas or optical radiation that is not necessarily far detuned. The
interactions introduced by $\hat{H}_S$ add dissipative and incoherent
processes to the reservoirs so that they act as semiclassical systems
equivalent to metallic electrodes. These processes are assumed to destroy
coherence and to relax the fermions to the ground state on a time scale
shorter than $\hbar/J_0$, making it possible to attribute a Fermi
distribution with well-defined temperature and chemical potential to the
reservoirs. An explicit scheme suggested in Ref.~\cite{Griessner-PRL-2006}
achieves this aim through a combination of coherent laser excitations and
dissipation into an ambient superfluid. First, fermions with high momentum
are transferred into the first excited band of the optical lattice via a
Raman process. Subsequently, the excited states decay into the lowest Bloch
band due to emission of phonons into the superfluid. An iteration of this
procedure results in a stable Fermi distribution of the atoms in the
reservoirs.

To apply the Landauer formalism modified to account for the finite size of
the reservoirs, we introduce the number operators $\hat{N}_\alpha$, with
$\alpha = L$ or $R$, measuring the number of fermions in the reservoirs and
the expectation values $\av{\hat{N}_\alpha}$. We specify the state of the
system by the average particle number $\av{\hat{N}_\alpha}$ and the current
$\partial_t\av{\hat{N}_\alpha}$ through the central part. To obtain the
current in the Landauer formalism we treat the connecting optical lattice as
a scattering potential with the energy-dependent transmission $\tr(\va)$.
Hence the average current $\partial_t\av{\hat{N}_R}= -
\partial_t\av{\hat{N}_L}$ is the sum of all possible scattering transfers
between the two reservoirs
\begin{equation}\label{landauer}
    \partial_t\av{\hat{N}_R(t)} = \int\frac{\dd\va}{2\pi\hbar}\,\tr(\va)\left[f_L(\va,t) - f_R(\va,t)\right]\,.
\end{equation}
The Fermi functions of the reservoirs are
$f_{\alpha}(\va,t)=[\ee^{(\va-\mu_\alpha)/k_B T}+1]^{-1}$, with the
Boltzmann constant $k_B$, the temperature $T$, and the time-dependent
chemical potential $\mu_\alpha(t)$. Note that Eq.~(\ref{landauer}) is
approximately valid provided that $f_\alpha(\va,t)$, or equivalently
$\mu_\alpha(t)$, varies slowly on the microscopic time scale $\hbar/J_0$.
This can readily be achieved by either increasing the size of the reservoirs
$M$ or decreasing the transmission through the central part~$C$.

The chemical potential $\mu_\alpha$ is related to the particle number
$\av{\hat{N}_\alpha}$ since the reservoirs are finite and the total number
of fermions in the system is fixed. The implicit relation between
$\mu_\alpha$ and the particle number $\av{\hat{N}_\alpha}$ is
\begin{equation}\label{number}
    \av{\hat{N}_\alpha} = M\int\dd\va\,\rho(\va)f_\alpha(\va)\,,
\end{equation}
with the density of states of the reservoir given by $\rho(\va) =
1/\pi\sqrt{(2J_0)^2-(\va-2J_0)^2}$ for $0\leq\va\leq 4J_0$ and zero
otherwise. Upon solving Eq.~(\ref{number}) for $\mu_\alpha(t)$, either
analytically or numerically, one finds the chemical potential
$\mu_\alpha(t)$ as a function of $\av{\hat{N}_\alpha(t)}$. Thus, as a
consequence of the finite-size reservoirs the Landauer formula in
Eq.~(\ref{landauer}) becomes a closed (integro-differential) equation for
$\av{\hat{N}_\alpha(t)}$.

The transmission $\mathcal{T}(\va)$ is efficiently determined by use of the
NEGF approach~\cite{Caroli-JPC-1971,Cuevas-2010}. In terms of Green's
functions we have
\begin{equation}\label{transmi}
    \tr(\va) = |G_{1,m}(\va)|^2\,\Gamma_L(\va)\Gamma_R(\va)\,,
\end{equation}
where $G_{i,j}(\va)$ is the full retarded Green's function for the central
part and $\Gamma_\alpha(\va) = 2\pi J_\alpha^2\rho_\alpha(\va)$ describes
the coupling to the reservoirs. Here, $\rho_L(\va)$  and $\rho_R(\va)$ are,
respectively, the local density of states at sites $0$ and $m+1$ (see
Fig.~\ref{scheme}). Consequently $\mathcal{T}(\va)$ encodes the coherent
evolution of the central part, governed by $\hat{H}_C$, as well as the
coupling to the reservoirs.

To calculate the full Green's function $G_{i,j}(\va)$ from $\hat{H}_C$ we
start with the bare Green's function $g_{i,j}(\va)$ whose components obey
the equation
\begin{equation}\nonumber
  (\va - \va_i)g_{i,j} + \sum_k J_{i,k}g_{k,j} = \delta_{i,j}
\end{equation}
or equivalently $g(\va) = 1/(\va - H_C)$, with both the Green's function and
$H_C$ in matrix notation. The bare Green's function $g_{i,j}(\va)$ with
poles at the energy levels of $H_C$ would be sufficient to determine the
transmission if the coherent region $C$ were coupled weakly to the
reservoirs. However, the full propagation of a fermion between the sites $i$
and $j$ includes excursions into the reservoirs due to the coupling, which
leads to the broadening $\Gamma_\alpha\sim J_\alpha^2/J_0$ of the energy
levels. More precisely, the excursions result in corrections to
$g_{i,j}(\va)$ in terms of self-energies $\Sigma_L = J^2_L \tilde{g}_{0,0}$
and $\Sigma_R = J^2_R \tilde{g}_{m+1,m+1}$, where $\tilde{g}_{i,j}(\va)$
denote the Green's functions of the reservoirs. Including the coupling to
the reservoirs to all orders we arrive at the Dyson equation for the full
Green's function
\begin{equation}\label{dyson}
  G_{i,j} = g_{i,j} + g_{i,1}\Sigma_L G_{1,j} + g_{i,m}\Sigma_R G_{m,j}\,.
\end{equation}
The relevant matrix element for the transmission $G_{1,m}$ is then obtained
from Eq.~(\ref{dyson}) by solving a set of simultaneous equations for
$G_{1,1}$, $G_{m,m}$, $G_{1,m}$ and $G_{m,1}$. As a result one finds
$G_{1,m} = g_{1,m}/D$ with~\cite{Caroli-JPC-1971}
\begin{equation}\nonumber
    D = (1 - \Sigma_L g_{1,1})(1 - \Sigma_R g_{m,m}) - \Sigma_L\Sigma_R g^2_{1,m}
\end{equation}
assuming that $g_{1,m} = g_{m,1}$.

For our specific setup we treat the reservoirs as semi-infinite optical
lattices, for which the Green's function $\tilde{g}_{i,j}$  at the end site
$0$ reads~\cite{Cuevas-2010}
\begin{equation}\nonumber
 \tilde{g}_{0,0}(\va) = \left[(\va-2J_0) - \ii\sqrt{(2J_0)^2-(\va-2J_0)^2}\right]/2J_0^2
\end{equation}
and $\tilde{g}_{m+1,m+1} = \tilde{g}_{0,0}$; the local density of states is
given by $\rho_\alpha(\va) = -(1/\pi)\mathrm{Im}\tilde{g}_{0,0}(\va)$. These
relations allow us to find explicit expressions for the couplings
$\Gamma_\alpha$ and the self-energies $\Sigma_\alpha$ in Eq.~(\ref{dyson}).

%%%%%%%%%%%%%%%%%%%%%%%%%%%%%%%%%%%%%%%%%%%%%%%%%%%%%%%%%%%%
%%%%%%%%%%%%%%%%%%%%%%%%%%%%%%%%%%%%%%%%%%%%%%%%%%%%%%%%%%%%
%%%%%%%%%%%%%%%%%%%%%%%%%%%%%%%%%%%%%%%%%%%%%%%%%%%%%%%%%%%%

\section{Fluctuations and Damping}

So far our analysis was restricted to the average fermion number
$\av{\hat{N}_\alpha(t)}$ and the current $\partial_t\av{\hat{N}_\alpha(t)}$,
which are found from Eqs.~(\ref{landauer}) and~(\ref{number}) for a given
transmission $\tr(\va)$ in Eq.~(\ref{transmi}). We now turn our attention to
quantum and thermal fluctuations present in the system. To simplify the
problem of determining the fluctuations we treat creation and damping
processes separately and, if possible, add their effects together.

%%%%%%%%%%%%%%%%%%%%%%%%%%%%%%%%%%%%%%%%%%%%%%%%%%%%%%%%%%%%
\begin{figure}[t]\vspace{5pt}
\begin{center}
  \hspace{-20pt}\includegraphics[width = 180pt]{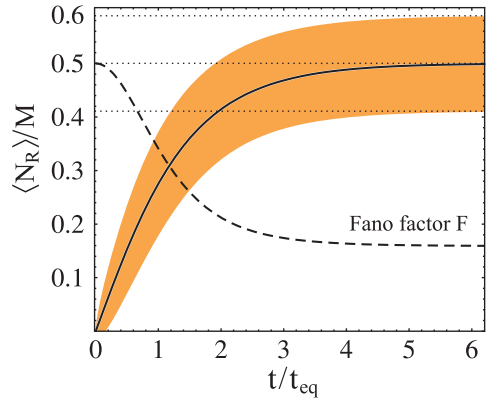}
  \caption{(Color online) Filling level and fluctuations in the right reservoir (initially empty) for a constant
  transmission~$\tr_0$. The average filling level $\av{N_R(t)}/M$ (solid line) increases
  with time $t/t_{eq}$ and saturates at the equilibrium value $1/2$.
  The standard deviation $\sqrt{\av{\delta N_R^2(t)}}/M$ from the average, due to thermal
  fluctuations, is indicated by the orange (gray) band. The Fano factor
  $F = \av{\delta N_R^2(t)}/\av{N_R(t)}$ (dashed line)
  decreases with time and approaches a constant value in the regime $t/t_{eq}\gg1$. The parameters
  are $k_B T/J_0 M = 1/10$ for the filling level and $k_B T/J_0 = 1$ for the Fano factor.}
  \label{thermal}
\end{center}
\end{figure}
%%%%%%%%%%%%%%%%%%%%%%%%%%%%%%%%%%%%%%%%%%%%%%%%%%%%%%%%%%%%

Fluctuations created during the evolution of the system can be found by
using the Levitov formula~\cite{Nazarov-2009}, which yields the full
counting statistics of thermal and quantum fluctuations provided
$\av{\hat{N}_\alpha(t)}$ and hence $\mu_\alpha(t)$ are known. However, we
limit our analysis to the most relevant statistical parameter, namely the
variance of the number of fermions $\av{\delta
N_\alpha^2}\equiv\av{(\hat{N}_\alpha - \av{\hat{N}_\alpha})^2}$. With
initially no fluctuations present we find for the variance $\av{\delta
N_\alpha^2(t)}$ after a time $t$ according to the Levitov formula
\begin{equation}\label{fluc}
\begin{split}
    \av{\delta N_\alpha^2(t)} = & \int_0^t\dd s \int\frac{\dd\va}{2\pi\hbar}\left\{\tr(\va)\bar{\tr}(\va)[f_L(\va,s) - f_R(\va,s)]^2\right.\\
    \vphantom{\int}+ &\left.\tr(\va)[f_L(\va,s)\bar{f}_L(\va,s) + f_R(\va,s)\bar{f}_R(\va,s)]\right\}\,,
\end{split}
\end{equation}
where we introduced $\bar{f}_\alpha(\va,t) = 1 - f_\alpha(\va,t)$ and
$\bar{\tr}(\va) = 1 - \tr(\va)$. At zero temperature, Eq.~(\ref{fluc})
describes the creation of quantum fluctuations caused by the probabilistic
nature of the particle transfer through the optical lattice. On the other
hand, at equilibrium between the reservoirs and for $\tr(\va)\equiv 1$ the
fluctuations are purely thermal.

Unlike in conventional mesoscopic systems with infinite-sized electrodes,
the fluctuations described by Eq.~(\ref{fluc}) are constantly damped out at
a rate $\gamma$. This intrinsic damping occurs because fluctuations in the
current immediately lead to fluctuations of the chemical potentials, which
drive the system back to the evolution according to the mean-field
description $\av{\hat{N}_\alpha(t)}$. To find an explicit expression for the
intrinsic damping $\gamma$ we determine the change in the chemical potential
$\delta\mu_\alpha$ caused by an excess of particles $\delta
N_\alpha\ll\av{\hat{N}_\alpha}$ with respect to $\av{\hat{N}_\alpha}$. From
Eq.~(\ref{number}) we obtain $\av{\hat{N}_\alpha} + \delta N_\alpha =
M\int\dd\va\rho(\va)f_\alpha(\va,\mu_\alpha+\delta\mu_\alpha)$, which to
lowest order in $\delta\mu_\alpha$ and $\delta N_\alpha$ results in the
linear dependence $\delta\mu_\alpha = \delta N_\alpha/M\rho_T(\mu_\alpha)$
with
\begin{equation}\label{rhotherm}
 \rho_T(\mu_\alpha) = \frac{1}{4k_B T}\int\dd\va\,\rho(\va)\,\mathrm{sech}^2\left(\frac{\va-\mu_\alpha}{2 k_B T}\right)\,.
\end{equation}
Note that $\rho_T(\mu_\alpha)$ reduces to $\rho(\mu_\alpha)$ in the case
$T=0$ and to $k_B T/4$ in the limit of infinite temperature. Similarly, we
expand the Landauer formula to lowest order in $\delta\mu_\alpha$ and
$\partial_t(\delta N_\alpha)$ and use the relation $\delta\mu_\alpha =
\delta N_\alpha/M\rho_T(\mu_\alpha)$ to find the time dependence for small
fluctuations $\partial_t(\delta N_\alpha) = -\gamma\,\delta N_\alpha$ with
the damping factor
\begin{equation}\label{damp}
    \gamma = \frac{1}{2\pi\hbar M}\left[\frac{\tr_T(\mu_L)}{\rho_T(\mu_L)}+\frac{\tr_T(\mu_R)}{\rho_T(\mu_R)}\right]\,,
\end{equation}
where $\tr_T(\mu_\alpha)$ is defined in the same way as $\rho_T(\mu_\alpha)$
in Eq.~(\ref{rhotherm}). The factor $\gamma$ depends on the filling level of
the reservoirs and is therefore time dependent in general. Importantly,
$\gamma$ is always positive and thus fluctuations are indeed damped out.

As a result, the fluctuations in the system depend not only on the
properties of the coherent region but also on the reservoirs via their
density of states $\rho(\va)$ evaluated at the filling level. Thus by
choosing the appropriate reservoirs it should be possible to perform
experiments in either the fluctuation-dominated or the mean-field regime.
This works particularly well for an optical lattice reservoir since its
density of states varies considerably over the entire bandwidth $4J_0$ so
that the intrinsic damping~$\gamma$ can be tuned over a wide range.

%%%%%%%%%%%%%%%%%%%%%%%%%%%%%%%%%%%%%%%%%%%%%%%%%%%%%%%%%%%%
%%%%%%%%%%%%%%%%%%%%%%%%%%%%%%%%%%%%%%%%%%%%%%%%%%%%%%%%%%%%
%%%%%%%%%%%%%%%%%%%%%%%%%%%%%%%%%%%%%%%%%%%%%%%%%%%%%%%%%%%%

%%%%%%%%%%%%%%%%%%%%%%%%%%%%%%%%%%%%%%%%%%%%%%%%%%%%%%%%%%%%
\begin{figure}[t]\vspace{5pt}
\begin{center}
  \hspace{-20pt}\includegraphics[width = 180pt]{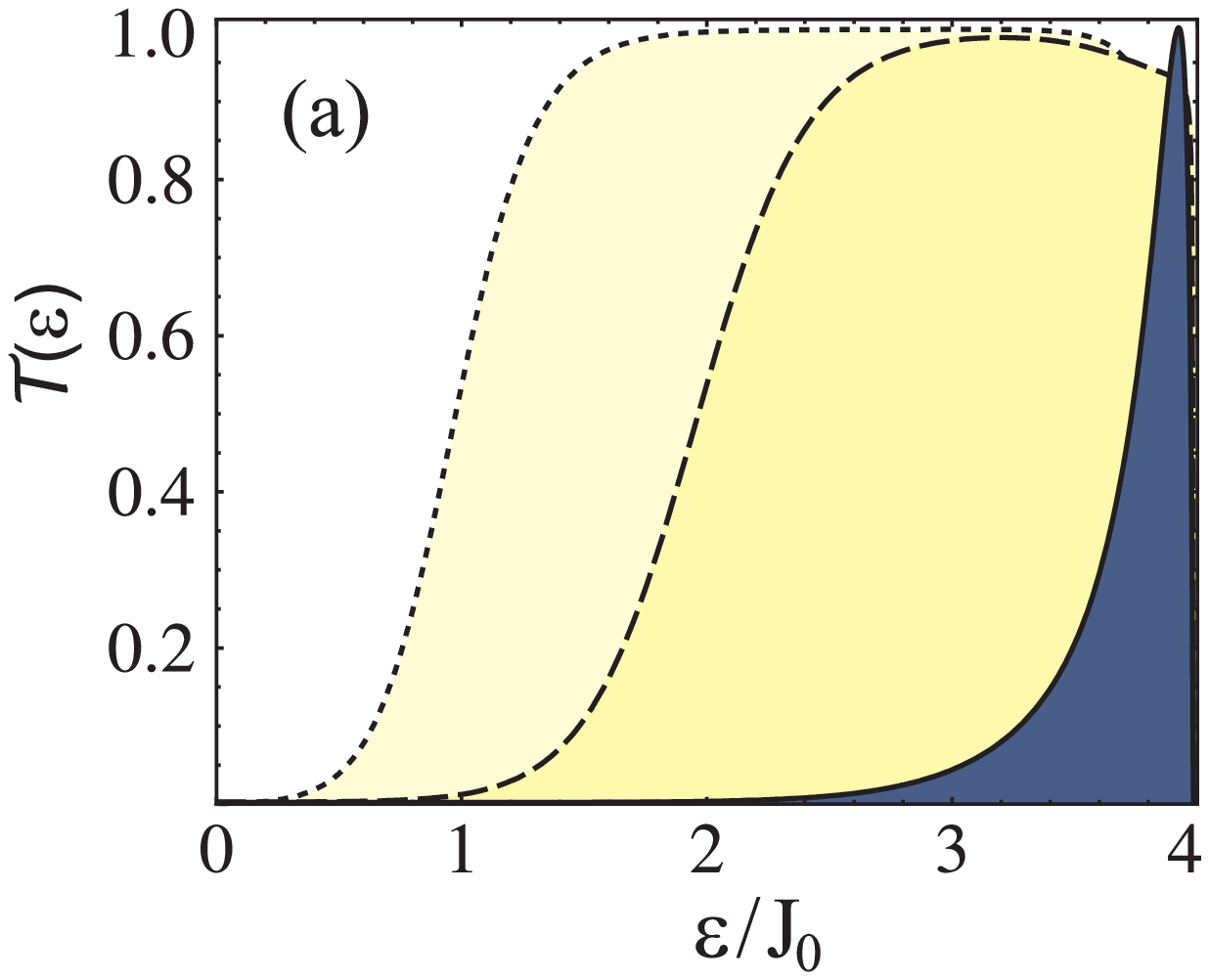}\vspace{15pt}\\
  \hspace{-20pt}\includegraphics[width = 180pt]{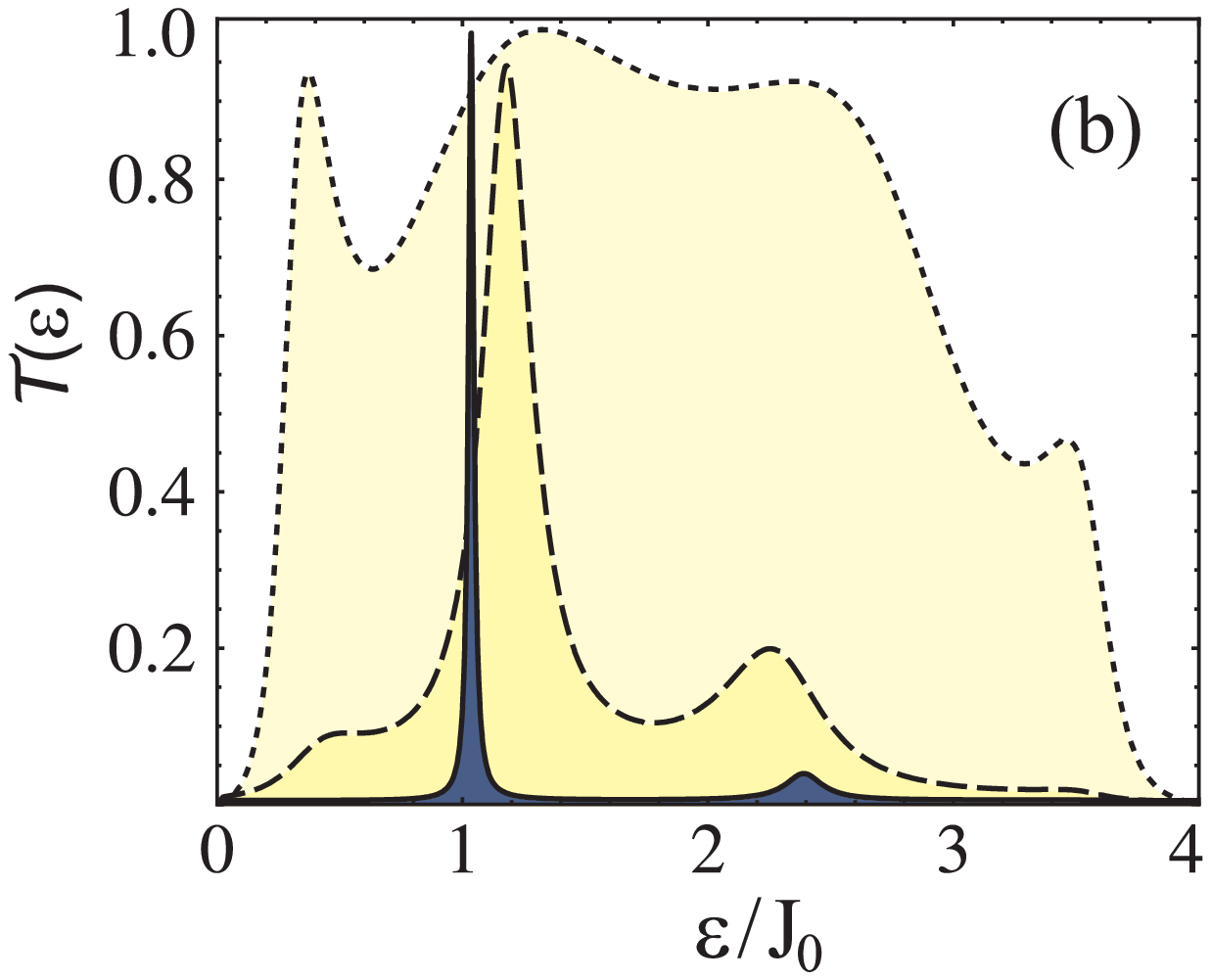}
  \caption{(Color online) The engineered transmission $\tr(\va)$
  as a function of the energy $\va/J_0$ for different modulations of an
  optical lattice with length $m=10$. (a)~A single centered beam with waist $\sigma = 2$
  and increasing depths $V/J_0 = {1,2,4}$ (dotted, dashed and full line)
  shifts the energies out of the reservoir band and reduces the
  transmission to a peak at the upper band limit. (b)~Two beams isolate a few
  central lattice sites and create a single resonant level at
  $\va/J_0\approx 1$ for sufficiently strong intensities.
  The beams are positioned at $\nu_1 = 3$, $\nu_2 = 8$ with waist $\sigma = 1$, depths $V_2 = {1,3,5}$
  (dotted, dashed and full line) and $V_1/V_2 = 1/2$.}\label{single}
\end{center}
\end{figure}
%%%%%%%%%%%%%%%%%%%%%%%%%%%%%%%%%%%%%%%%%%%%%%%%%%%%%%%%%%%%

\section{Constant Transmission}

To gain physical insight into the transport between reservoirs we apply our
general results to the important special case of constant transmission:
$\tr(\va)\equiv\tr_0$ for $0\leq\va\leq 4J_0$ and zero otherwise, where the
constant transmission $\tr_0$ takes values $0\leq\tr_0\leq1$. In this case
the Landauer formula for the current reduces to $\partial_t\av{\hat{N}_R} =
\tr_0\Delta\mu/2\pi\hbar$ and hence one only has to determine the dependence
of $\Delta\mu = \mu_L-\mu_R$ on the average particle number
$\av{\hat{N}_\alpha}$ and the temperature $T$.

For concreteness we consider the equilibration process in the
low-temperature regime $k_B T\ll J_0$, neglecting corrections of the order
$(k_B T/J_0)^2$. In this case the chemical potential difference is
$\Delta\mu = 4J_0\cos(\pi \av{\hat{N}_R}/M)$ assuming the initial conditions
$\av{\hat{N}_L} = M$ and $\av{\hat{N}_R} = 0$. Solving the Landauer equation
we obtain the evolution of the filling level
\begin{equation}\nonumber
    \av{\hat{N}_R(t)} = CU\arctan\left[\sinh(t/t_{eq})\right]
\end{equation}
and the current through the central region $\partial_t\av{\hat{N}_R(t)} =
(U/R)\,\mathrm{sech}(t/t_{eq})$. The equilibration time scale $t_{eq} = RC$
satisfies $t_{eq}\gg\hbar/J_0$ in the relevant parameter regime as required
for Eq.~(\ref{landauer}) to be valid. Here, analogous to a classical RC
circuit, we introduced the resistance $R = 2\pi\hbar/\tr_0$, the capacitance
$C = M/4\pi J_0$, and initial capacitor voltage $U = 4J_0$.

We next determine the evolution of the fluctuations during the equilibration
process. For simplicity we assume from now on that $\tr_0$ is close to
unity, i.e., $\tr_0\approx1$, so that predominantly thermal fluctuations
with $\av{\delta N_\alpha^2(t)} = 2t k_B T/R$ are created during the
equilibration process, as can be found from the Levitov formula in
Eq.~(\ref{fluc}). Since correlations of thermal fluctuations decay fast on
the time scale $t_{eq}$, the evolution of the deviations $\delta N_R$ from
$\av{\hat{N}_R(t)}$ can be expressed in the form of a Langevin equation
\begin{equation}\label{langevin}\nonumber
    \partial_t(\delta N_R) = -\gamma(t)\delta N_R + \xi(t)\,.
\end{equation}
The time-dependent damping factor resulting from Eq.~(\ref{damp}) reads
$\gamma(t) = \sin[\pi \av{\hat{N}_R(t)}/M]/RC$ and thus damping increases as
the system approaches equilibrium. The fluctuations are represented by the
stochastic force $\xi(t)$ and satisfy the condition
$\langle\xi(t)\xi(t^\prime)\rangle = (2 k_B T/R)\,\delta(t-t^\prime)$. Using
standard techniques for stochastic problems~\cite{Gardiner-2004} we find the
evolution of the fluctuations
\begin{equation}\nonumber
    \av{\delta N_R^2(t)} = Ck_B T
\left[\mathrm{sech}^2(t/t_{eq})\,t/t_{eq}+\tanh(t/t_{eq})\right]\,.
\end{equation}
We see that the fluctuations increase linearly in the regime
$t/t_{eq}\ll1$, where damping is weak according to Eq.~(\ref{damp}). On the
other hand, in the limit $t/t_{eq}\gg1$ the fluctuations converge to the
constant value $C k_B T$, which results from the competition between thermal
fluctuations and the intrinsic damping $\gamma = 1/RC$ at equilibrium.

Figure~\ref{thermal} shows the time evolution of the average filling level
$\av{\hat{N}_R(t)}/M$ and the fluctuations in the right reservoir. The
filling level increases linearly in the regime $t/t_{eq}\ll1$ and saturates
at the equilibrium value $\av{\hat{N}_R} = \av{\hat{N}_L} = M/2$. Thermal
fluctuations around the average are limited by damping; however, they
indicate significant shot-to-shot deviations from the average filling level.
Of particular interest is the Fano factor defined by $F = \av{\delta
N_R^2(t)}/\av{\hat{N}_R(t)}$, which is independent of the size of the
reservoirs $M$ and gives a direct measure of the temperature of the system.
The Fano factor converges to the value $F = k_B T/2\pi J_0$ as the system
equilibrates, i.e., in the limit $t/t_{eq}\gg1$. This result even holds for
arbitrary constant transmissions $\tr_0$ as quantum fluctuations are damped
out in this limit.

An elementary experimental configuration with constant transmission, namely
$\tr_0 \approx 1$, consists of a single homogeneous optical lattice
partitioned into a sufficiently long coherent part $C$ and the reservoirs
$L$ and $R$. The corresponding hopping parameters and couplings are $J_{i,j}
= J_0$ and $J_\alpha = J_0$, respectively. In fact, the strong couplings
$J_\alpha$ lead to significant broadening of the cosine-distributed energy
levels of the coherent part \footnote{The energy levels $E_\ell$ of the
uncoupled coherent part $C$ with constant hopping $J_0$ are given by $E_\ell
= 2J_0-2J_0\cos\left[\pi \ell/(m+1)\right]$ with $\ell = 1,2,\ldots,m$.}. As
a consequence, the broadened energy levels merge together in the regime
$J_\alpha\gg 2\pi J_0/m$, which results in an approximately constant
transmission.

%%%%%%%%%%%%%%%%%%%%%%%%%%%%%%%%%%%%%%%%%%%%%%%%%%%%%%%%%%%%
%%%%%%%%%%%%%%%%%%%%%%%%%%%%%%%%%%%%%%%%%%%%%%%%%%%%%%%%%%%%
%%%%%%%%%%%%%%%%%%%%%%%%%%%%%%%%%%%%%%%%%%%%%%%%%%%%%%%%%%%%

%%%%%%%%%%%%%%%%%%%%%%%%%%%%%%%%%%%%%%%%%%%%%%%%%%%%%%%%%%%%
\begin{figure}[t]\vspace{5pt}
\begin{center}
  \hspace{-20pt}\includegraphics[width = 180pt]{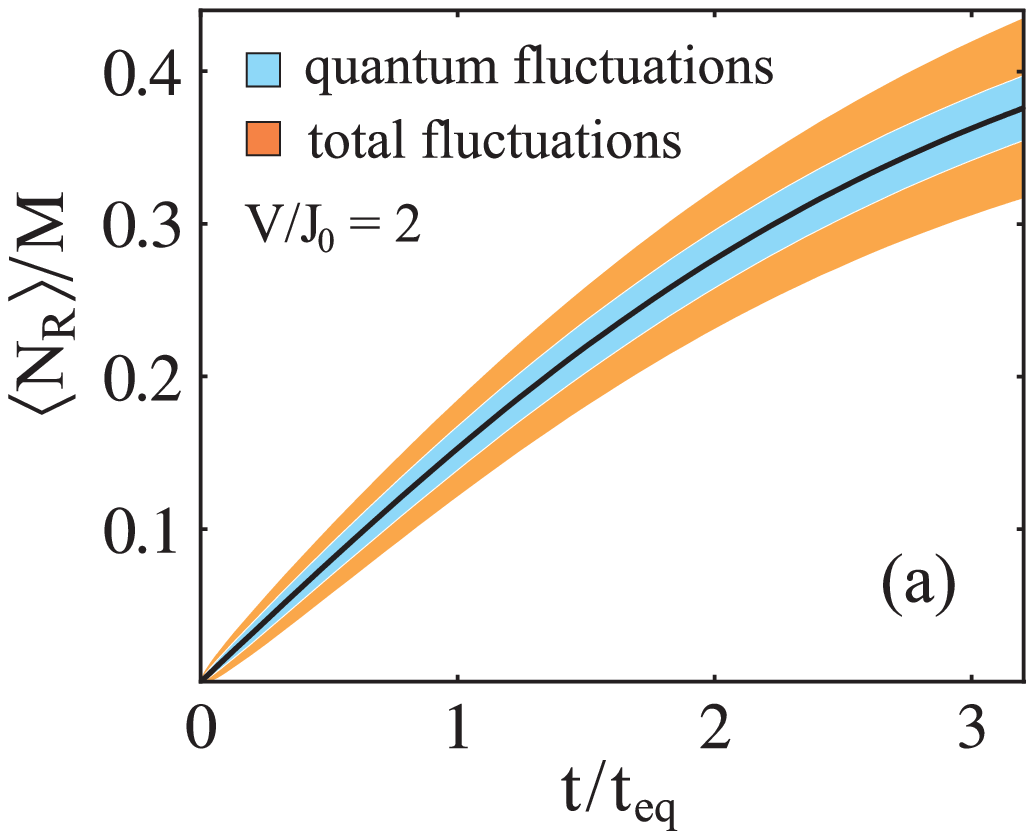}\vspace{15pt}\\
  \hspace{-24pt}\includegraphics[width = 184pt]{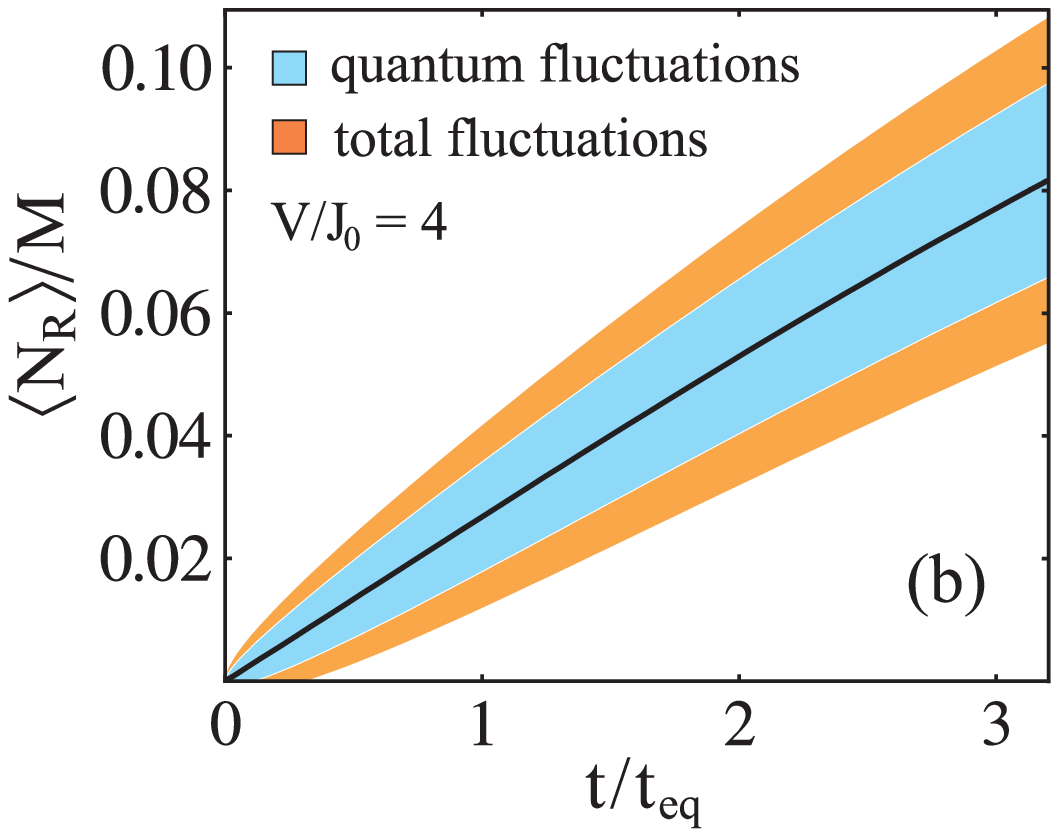}
  \caption{(Color online) Filling level and fluctuations in the right reservoir (initially empty) for a
  lattice modulated by a single beam. (a)~The average filling level
  $\av{N_R(t)}/M$ (solid line) as a function time $t/t_{eq}$ (with $t_{eq}=\hbar M/2J_0$) for
  the modulation strength $V/J_0 = 2$. Different
  bands indicate the standard deviation $\sqrt{\av{\delta N_R^2(t)}}/M$ due to
  quantum fluctuations (blue or light gray) and total fluctuations, i.e., quantum plus thermal (orange or dark gray).
  (b)~The same quantities for
  a stronger modulation strength $V/J_0 = 4$. The lower transmission results in a smaller current between
  the reservoirs and enhanced quantum fluctuations in comparison to (a). In both plots damping
  has not been taken into account and the parameters are $k_B T/J_0 = 1$, $m=10$ and $M=50$.}\label{noise}
\end{center}
\end{figure}
%%%%%%%%%%%%%%%%%%%%%%%%%%%%%%%%%%%%%%%%%%%%%%%%%%%%%%%%%%%%

\section{Transmission Engineering}

The usefulness of the NEGF-Landauer formalism is most evident if we want to
calculate the current through an engineered optical lattice with tailored
parameters $J_{i,j}$ and $\va_j$. Since the optical lattice potentials for
the reservoirs and the central region would most likely be produced by the
same counterpropagating laser beams we set $J_{i,j} = J_\alpha = J_0$ and
focus on modified on-site energies $\va_j$. A possible experimental
configuration with tailored on-site energies $\va_j$ involves one (or
several) laser beams crossing the central region perpendicular to the
optical lattice. The optical potential caused by a single beam centered at
position $\nu$ shifts the energies as
\begin{equation}\nonumber
\va_j = V\exp[-(j-\nu)^2/\sigma^2]\,,
\end{equation}
with the potential strength $V$ and width $\sigma$ measured in units of
lattice spacings. Depending on the detuning, $V$ may take positive or
negative values~\cite{Grimm-ADV-2000}. We emphasize that unlike the
scheme proposed in Ref.~\cite{Pepino-PRL-2009} such a configuration does not
require site-by-site control of the optical lattice, neither of the on-site
energies $\va_j$ nor the hopping parameters $J_{i,j}$ or $J_\alpha$.

A specific setup consists of a single laser beam, centered at $\nu\approx
m/2$ and with beam waist $\sigma\sim m/4$, acting as a $V$-dependent switch
for the fermion current. The potential shifts the energies $\va_j$ out of
the reservoir band and thus reduces the transmission significantly, as shown
in Fig.~\ref{single}(a). This configuration provides the possibility to
study the dependence of the fluctuations on the transmission $\tr(\va)$ and
the temperature $T$ of the fermions: If we choose the initial conditions
$\av{\hat{N}_L} = M$, $\av{\hat{N}_R} = 0$ and stay far from equilibrium
then according to Eq.~(\ref{damp}) damping is negligible. As a consequence,
thermal and quantum fluctuations lead to significant deviations of the
filling level from $\av{\hat{N}_R(t)}$, which are detectable by counting the
actual number of fermions in the right reservoir. Figure~\ref{noise} shows
the average particle number $\av{\hat{N}_R(t)}$ and the expected
fluctuations for modulations of the optical lattice with two different
potential strengths~$V$. At zero temperature, only quantum fluctuations
caused by the limited transmission contribute, whereas at finite temperature
fluctuations are further increased. By comparing Figs.~\ref{noise}(a) and
\ref{noise}(b) we see that quantum fluctuations become more important for
reduced transmissions, i.e., for stronger potential strengths $V$.

A paradigmatic system in the context of mesoscopic physics is the
single-level model~\cite{Cuevas-2010}, or in the case of interacting
fermions the Anderson impurity model~\cite{Anderson-PR-1961}. These models
may be used, e.g., to study the Kondo effect or to describe transport
through a single quantum dot. Realization of a single-level model can be
achieved by means of two laser beams with different detuning leading to
energy shifts
\begin{equation}\nonumber
    \va_j = V_1\exp[-(j-\nu_1)^2/\sigma^2] - V_2\exp[-(j-\nu_2)^2/\sigma^2]\,.
\end{equation}
If the beams are separated with $\nu_1-\nu_2\sim m/2$ and narrow $\sigma\sim
m/4$ then for sufficiently strong potentials $V_1, V_2>0$ the transmission
exhibits a single peak, as shown in Fig.~\ref{single}(b). The position of
the peak depends on the ratio $V_1/V_2$ and the strength of the potentials
determines the width of the peak, i.e., the effective coupling to the
reservoirs.

The emergence of the single level can be understood in the energy band
picture: The first beam shifts the unmodulated band of width $4J_0$ upward,
while the second beam shifts the band downward. As a consequence, a small
region between the beams is isolated from the reservoirs and acts as a
single energy level. The effective couplings $\Gamma_\alpha$ of the level to
the reservoirs, or equivalently the width of the single level, is readily
controllable by the strengths of the beams $V_1$ and $V_2$. This makes it
possible to access the weak-coupling regime considered in
Ref.~\cite{Pepino-PRL-2009} as well as the strong-coupling regime without
the requirement of specific control of the hopping parameters $J_\alpha$.

%%%%%%%%%%%%%%%%%%%%%%%%%%%%%%%%%%%%%%%%%%%%%%%%%%%%%%%%%%%%
%%%%%%%%%%%%%%%%%%%%%%%%%%%%%%%%%%%%%%%%%%%%%%%%%%%%%%%%%%%%
%%%%%%%%%%%%%%%%%%%%%%%%%%%%%%%%%%%%%%%%%%%%%%%%%%%%%%%%%%%%

\section{Effect of Interactions}

Let us now discuss the effect of weak interactions between the fermions on
the basis of the single-level model. If we consider a spin-balanced mixture
of fermions in two different internal states, denoted by up and down, then
the total Hamiltonian in Eq.~(\ref{ham}) is augmented by the interaction
term $\hat{H}_U = U\sum_j\hat{n}_j^\uparrow\hat{n}_j^\downarrow$. Here, $U$
is the either positive or negative interaction strength between the fermions
and $\hat{n}_j^\uparrow$ ($\hat{n}_j^\downarrow$) is the occupation number
operator for the up (down) states. For our analysis we use the mean-field
approximation
$\hat{n}_j^\uparrow\hat{n}_j^\downarrow\approx\hat{n}_j^\uparrow\av{\hat{n}_j^\downarrow}
+\av{\hat{n}_j^\uparrow}\hat{n}_j^\downarrow -
\av{\hat{n}_j^\uparrow}\av{\hat{n}_j^\downarrow}$ which we expect to be
valid for sufficiently small reservoirs and low filling
levels~\cite{Linden-PRB-1990}. The chemical potentials of the reservoirs
$\mu_\alpha$ are changed accordingly due to the interactions; e.g., the
chemical potential for spin-up states is given by $\mu_\alpha^\uparrow =
\mu_\alpha + U\av{\hat{n}^\downarrow}$. In the following we will however
neglect these trivial changes.

%%%%%%%%%%%%%%%%%%%%%%%%%%%%%%%%%%%%%%%%%%%%%%%%%%%%%%%%%%%%
\begin{figure}[t]\vspace{5pt}
\begin{center}
  \includegraphics[width = 190pt]{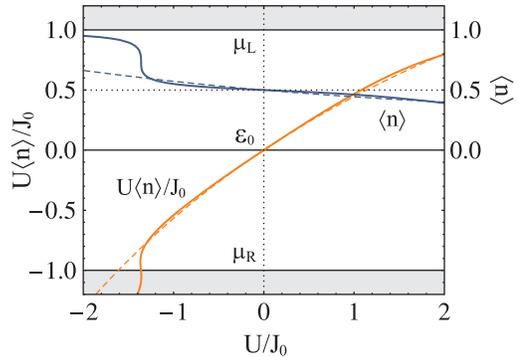}
  \caption{(Color online) The effect of interactions on a single level
  with original ($U=0$) position $\va_0 = 0$ and the relative chemical
  potentials $\mu_L/J_0 = 1$ and  $\mu_R/J_0 = -1$. The occupation $\av{n}$ (blue or dark gray)
  and the energy shift $U\av{n}/J_0$ (orange or light gray) as a function of the interaction~$U/J_0$
  are plotted for the couplings $\Gamma/J_0 = 0.1$ (solid) and $\Gamma/J_0 = 1$ (dashed).
  For $U>0$ the level is depleted and shifted toward $\mu_L$. For $U<0$ the level
  is almost completely occupied and shifted below $\mu_R$ for sufficiently strong interactions.
  The current is blocked if the single level at $\va_0+U\av{n}$ leaves the energy window $[\mu_L,\mu_R]$.}
  \label{occupation}
\end{center}
\end{figure}
%%%%%%%%%%%%%%%%%%%%%%%%%%%%%%%%%%%%%%%%%%%%%%%%%%%%%%%%%%%%

The effect of interactions on the single level is to shift its original
($U=0$) position $\va_0$ by the interaction energy; e.g., for spin-up
fermions the level position is $\va_0 + U\av{\hat{n}^\downarrow}$. The
corresponding Green's function reads
\begin{equation}
    G^\uparrow(\va) = 1/[\va-\va_0 - U\av{\hat{n}^\downarrow} +
\ii\Gamma_L(\va)+ \ii\Gamma_R(\va)]
\end{equation}
with the effective couplings $\Gamma_\alpha$. The occupation of the single
level is given by
\begin{equation}\label{occ}
    \av{\hat{n}^\uparrow} = \int\frac{\dd\va}{2\pi}\,|G^\uparrow(\va)|^2[\Gamma_L(\va)f_L(\va) + \Gamma_R(\va)f_R(\va)]\,,
\end{equation}
with the time dependence of the Fermi functions omitted. In the case of a
spin-balanced mixture with $\av{\hat{n}^\uparrow} = \av{\hat{n}^\downarrow}$
the Green's function $G^\uparrow(\va)$ and the occupation
$\av{\hat{n}^\uparrow}$ can be determined self-consistently to obtain the
transmission $\tr(\va)\propto|G^\uparrow(\va)|^2$, which is the same for
both internal states. In principle, the average particle number
$\av{\hat{N}_\alpha(t)}$ and the current $\partial_t\av{\hat{N}_\alpha(t)}$
are then evaluated as for noninteracting fermions.

The effect of interactions on the transmission can be qualitatively
understood in the wide-band limit with constant couplings $\Gamma_\alpha =
\Gamma$. Inserting $G^\uparrow(\va)$ into Eq.~(\ref{occ}) then yields the
self-consistent equation for the occupation~\cite{Anderson-PR-1961}
\begin{equation}\label{selfocc}
    \av{\hat{n}} = \frac{1}{2} + \frac{2}{\pi}\sum_\alpha\arctan\left(\frac{\mu_\alpha-\va_0-U\av{\hat{n}}}{2\Gamma}\right)\,,
\end{equation}
where the average occupation number applies to both spin states;
i.e.,~$\av{\hat{n}} = \av{\hat{n}^\uparrow} = \av{\hat{n}^\downarrow}$.
Figure~\ref{occupation} shows the average occupation $\av{\hat{n}}$ of the
single level and the energy shift $U\av{\hat{n}}$ as a function of the
interaction strength~$U$ according to Eq.~(\ref{selfocc}). For repulsive
interactions $U>0$ we observe a depletion of the single level and a shift to
higher energies. This shift is bounded by $\mu_L$ in the limit of vanishing
coupling $\Gamma$, but takes values larger than $\mu_L$ in the case of
finite $\Gamma$. For attractive interactions $U<0$ the occupation
$\av{\hat{n}}$ increases and makes an abrupt transition to
$\av{\hat{n}}\approx 1$ accompanied by a shift of the level below $\mu_R$.
In both cases the single level at $\va_0+U\av{\hat{n}}$ eventually leaves
the energy window between $\mu_L$ and $\mu_R$, and hence the current through
the level is strongly suppressed. Thus interactions offer an alternative
approach to control the current through the lattice.

%%%%%%%%%%%%%%%%%%%%%%%%%%%%%%%%%%%%%%%%%%%%%%%%%%%%%%%%%%%%
%%%%%%%%%%%%%%%%%%%%%%%%%%%%%%%%%%%%%%%%%%%%%%%%%%%%%%%%%%%%
%%%%%%%%%%%%%%%%%%%%%%%%%%%%%%%%%%%%%%%%%%%%%%%%%%%%%%%%%%%%

\section{Conclusions}

Using the mesoscopic NEGF-Landauer approach we have analyzed nonequilibrium
transport of fermions through an engineered optical lattice for arbitrarily
strong coupling to two reservoirs at finite temperatures. We have
characterized the full equilibration process by calculating the accumulated
number of atoms in the finite reservoirs, which is a directly accessible
quantity in experiments. Considering experimentally relevant system
parameters we found that the reservoirs equilibrate on time scales
comparable to the duration of typical ultracold atom experiments. Our
systematic analysis of created and damped fluctuations in the finite system
revealed that the mean-field description gives an incomplete picture of
fermion transport since significant shot-to-shot variations from the average
current, partly due to thermal effects, are to be expected. This is directly
relevant to the emulation of semiconductor electronic circuits, where
preferably single-shot measurements are required to determine the
current~\cite{Pepino-PRA-2010}.

The study of fluctuations around the average current revealed additional
information about the processes taking place in the system: We found that
thermal fluctuations build up on the time scale of the equilibration process
until they reach a constant value proportional to the temperature of the
Fermi gas. As an aside, we note that thermal fluctuation between
equilibrated reservoirs may therefore be used for thermometry of the system.
On a more fundamental level, we saw that a decrease in the current due to a
lower transmission of the coherent region is necessarily accompanied by
higher quantum fluctuations. This correlation allows the experimenter, e.g.,
to distinguish between changes in either the chemical potentials or the
transmission as the cause of a reduced current.

We have shown that modulations of a homogeneous lattice potential can be
used not only to reduce the equilibration current, but also to realize a
single-level model with full control over the position and the coupling of
the level. This setup requires neither additional impurity atoms nor
site-by-site manipulations of the optical lattice. Moreover, advanced
experimental techniques for producing tailor-made optical potentials, by
employing either acousto-optical deflectors~\cite{Zimmermann-NJP-2011} or
holographic mask techniques~\cite{Pasienski-OE-2008}, are expected to
further facilitate the creation of engineered optical lattices. Finally, our
mean-field analysis of interaction effects revealed that even weak
interactions between the fermions suppress the current through a single
level, which can be exploited to control the current.

We conclude with the observation that our approach to nonequilibrium
transport between finite reservoirs may be applied to similar ultracold
atomic setups~\cite{Gadway-AX-2011} or, more generally, to mesoscopic
systems such as electrons on liquid helium~\cite{Rees-PRL-2011}. Possible
extensions of this work include transport of bosonic atoms, similar to the
analysis in Refs.~\cite{Micheli-PRA-2006,Chien-AX-2011}, quantum pumping
between reservoirs using time-dependent modulations of the optical
lattice~\cite{Das-PRL-2009}, and the effect of interactions between fermions
on quantum fluctuations~\cite{Kambly-PRB-2011}.

%%%%%%%%%%%%%%%%%%%%%%%%%%%%%%%%%%%%%%%%%%%%%%%%%%%%%%%%%%%%
%%%%%%%%%%%%%%%%%%%%%%%%%%%%%%%%%%%%%%%%%%%%%%%%%%%%%%%%%%%%
%%%%%%%%%%%%%%%%%%%%%%%%%%%%%%%%%%%%%%%%%%%%%%%%%%%%%%%%%%%%

\acknowledgments

M.B. thanks Leticia Tarruell for valuable discussions on experimental
aspects and Stephen R. Clark for providing clarifying numerical results on
nonequilibrium transport. M.B. and W.B. acknowledge financial support from
the German Research Foundation (DFG) through SFB 767 and the Swiss National
Science Foundation (SNSF) through Project PBSKP2/130366.

\bibliography{trans}

\end{document}